\documentclass[10pt,conference]{IEEEtran}
\usepackage[T1]{fontenc}
\IEEEoverridecommandlockouts

\makeatletter
\def\ps@headings{%
\def\@oddhead{\mbox{}\scriptsize\rightmark \hfil \thepage}%
\def\@evenhead{\scriptsize\thepage \hfil \leftmark\mbox{}}%
\def\@oddfoot{}%
\def\@evenfoot{}}
\makeatother
\usepackage{subfigure}
\usepackage{colortbl}
\usepackage{bm}
\usepackage{graphicx}
\usepackage{epstopdf}

\usepackage{graphicx}  
\usepackage{url}       

\usepackage{amsmath}   
\usepackage{cite}
\usepackage{extarrows}
\usepackage{amsfonts,amssymb}

\usepackage{stfloats}

\usepackage{cases}
\usepackage{algorithm}
\usepackage{multirow}
\usepackage{algorithmic}

\usepackage{amsmath}
\allowdisplaybreaks[4]

\newcommand{\ls}[1]
    {\dimen0=\fontdimen6\the\font
     \lineskip=#1\dimen0
     \advance\lineskip.5\fontdimen5\the\font
     \advance\lineskip-\dimen0
     \lineskiplimit=.9\lineskip
     \baselineskip=\lineskip
     \advance\baselineskip\dimen0
     \normallineskip\lineskip
     \normallineskiplimit\lineskiplimit
     \normalbaselineskip\baselineskip
     \ignorespaces
    }

\graphicspath{{Matlab/}{Visio/}}

\IEEEoverridecommandlockouts\IEEEpubid{\makebox[\columnwidth]{979-8-3503-1090-0/23/\$31.00~\copyright~2023 IEEE \hfill} \hspace{\columnsep}\makebox[\columnwidth]{ }}
\begin{document}
\pagestyle{empty}
\title{Performance Analysis and Blocklength Minimization of Uplink RSMA for Short \\Packet Transmissions in URLLC}

\author{\IEEEauthorblockN{Yixin Zhang$^{\dagger}$, Wenchi Cheng$^{\dagger}$, Jingqing Wang$^{\dagger}$, and Wei Zhang$^{\ddagger}$}~\\[0.2cm]
\vspace{-8pt}

\IEEEauthorblockA{$^{\dagger}$State Key Laboratory of Integrated
Services Networks, Xidian University, Xi'an, China\\
$^{\ddagger}$School of Electrical Engineering and Telecommunications, The University of New South Wales, Sydney, Australia\\
E-mail: \{\emph{yixinzhang@stu.xidian.edu.cn}, \emph{wccheng@xidian.edu.cn}, \emph{jqwangxd@xidian.edu.cn}, \emph{w.zhang@unsw.edu.au}\}}

\vspace{-23pt}

}

\maketitle

\begin{abstract}
Rate splitting multiple access (RSMA) is one of the promising techniques for ultra-reliable and low-latency communications (URLLC) with stringent requirements on delay and reliability of multiple access. To fully explore the delay performance enhancement brought by uplink RSMA to URLLC, in this paper, we evaluate the performance of two-user uplink RSMA and propose the corresponding blocklength minimization problem. We analyze the impact of finite blocklength (FBL) code on the achievable rate region and the effective throughput of uplink RSMA. On this basis, we propose the problem of minimizing the blocklength for uplink RSMA with power allocation under constrained reliability and effective throughput. Then, we present an alternating optimization method to solve this non-convex problem. Simulation results show that different from the infinite blocklength (IBL) regime, the achievable rate region of the uplink RSMA is not always larger than that of uplink non-orthogonal multiple access (NOMA) in the FBL regime. But with the help of our proposed blocklength minimization scheme with power allocation, uplink RSMA can achieve the same achievable rate with a smaller blocklength compared to uplink NOMA, frequency division multiple access (FDMA), and time division multiple access (TDMA) in the FBL regime, showing the potential of uplink RSMA to achieve low delay without time sharing for URLLC.
\end{abstract}

\vspace{5pt}


\section{Introduction}
\IEEEPARstart{A} wide range of real-time applications and services, such as autonomous vehicles, Industrial Internet of Things (IIoT), and augmented reality/virtual reality (AR/VR), are emerging at a fast speed in the upcoming sixth generation (6G) wireless networks~\cite{6G_URLLC}. Ultra-reliable and low-latency communications (URLLC), as the core service in the fifth generation (5G) communication networks, aims to provide end-to-end (E2E) delay of less than 1 ms for 32-bit packet transmission while ensuring packet error probability of less than $10^{-5}$~\cite{38.913}. To meet the ultra-high quality of service (QoS) demands for various delay-sensitive services, more stringent requirement on delay is put forward as sub-millisecond level in 6G networks~\cite{sub_millisecond}. One of the key methods to meet the delay requirements of real-time applications is to use finite blocklength (FBL) code for short packet transmissions~\cite{Short_packet_1}. The traditional Shannon capacity, which is used in the infinite blocklength (IBL) for high-capacity-demanded services, is no longer applicable to short packet transmissions in URLLC. The authors derived a closed-form expression of the achievable rate and the decoding error probability in the FBL regime~\cite{FBL}. In addition, the design of future 6G networks also requires support for large-scale access to ensure stringent delay and reliability requirements from a large number of devices~\cite{MA_URLLC}. In the existing network architecture, delay timeout under multiple access (MA) conditions remains a difficult problem, where time-saving and reliable solutions are still very important for future networks. 

Recently, rate splitting multiple access (RSMA) has been proposed as a promising technology to enhance spectral efficiency (SE), energy efficiency (EE), coverage, QoS, user fairness, and reliability while entailing lower delay, feedback overhead, and complexity~\cite{RSMA_Survey}. RSMA relies on rate splitting (RS) and superposition coding (SC) at the transmitter, as well as successive interference cancellation (SIC) at the receiver. RSMA has been widely studied for downlink systems, showing the SE, EE, and delay enhancement of downlink RSMA~\cite{Downlink_RSMA_MIMO,Downlink_RSMA_FBL}. As for the uplink RSMA, it was first proposed in~\cite{RSMA_Region} for the single-input single-output (SISO) multiple access channel (MAC) to achieve every point of the Gaussian MAC capacity region without the need for time sharing and joint encoding-decoding among users. In this way, uplink RSMA can avoid high complexity and overhead, resulting in a relatively high rate and low delay. The throughput improvement of uplink RSMA has been studied in~\cite{Uplink_RSMA_FBL}. However, the delay performance with FBL code of uplink RSMA has not been studied, which means how to take advantage of RSMA in the URLLC system still needs to be addressed. To fully explore the delay performance improvement brought by RSMA to URLLC, the FBL and RSMA combined analysis and delay minimization with blocklength optimization are highly needed.

In order to solve this problem, in this paper we analyze the impact of FBL code on uplink RSMA to obtain the achievable rate region and the effective throughput of uplink RSMA in the FBL regime. Based on the above analysis, we propose the blocklength minimization problem with power allocation for uplink RSMA under constrained reliability and effective throughput. Then, we present an alternating optimization algorithm to solve this non-convex problem. Simulation results show that our proposed uplink RSMA-based blocklength minimization problem with power allocation can use a lower blocklength with a lower delay to achieve the same achievable rate compared to uplink non-orthogonal multiple access (NOMA), frequency division multiple access (FDMA), and time division multiple access (TDMA).

The rest of this paper is organized as follows. Section~\ref{Sec:System} introduces the two-user uplink RSMA system model. Section~\ref{Sec:FBL} analyzes the performance of uplink RSMA in the FBL regime. Section~\ref{Sec:Problem} presents the uplink RSMA-based blocklength minimization problem with power allocation and the corresponding algorithm. Section~\ref{Sec:Results} provides the numerical results. Finally, we conclude this paper in Section~\ref{Sec:Con}.

\section{System Model}\label{Sec:System}

\addtolength{\topmargin}{0.27in}

\begin{figure}[htbp]
	\centering\includegraphics[width=3.3in]{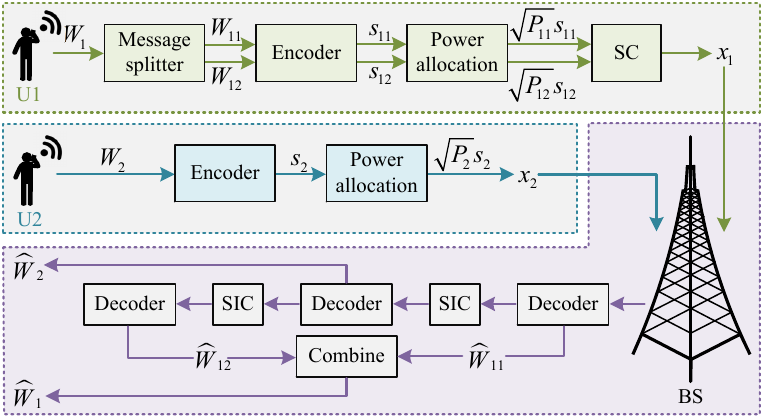}
	\caption{Two-user uplink RSMA system model.}\label{fig:System}
\end{figure}
As shown in Fig.~\ref{fig:System}, we consider a two-user uplink RSMA system consisting of one base station (BS) and two users U1 and U2. In this paper, 1-layer RSMA is adopted to serve U1 and U2. The message $W_1$ of U1 is split into two sub-messages $W_{11}$ and $W_{12}$, which can be interpreted as creating 2 virtual users at U1~\cite{RSMA_Survey}. The messages $W_{11}$ and $W_{12}$ are independently encoded into streams $s_{11}$ and $s_{12}$, which are then respectively allocated with certain powers $P_{11}$ and $P_{12}$. Thus, the transmit signal at U1 is given by
\begin{equation}
	x_1 = \sqrt{P_{11}}s_{11} + \sqrt{P_{12}}s_{12}.
\end{equation}
At U2, the message $W_2$ is directly encoded into $s_2$. By allocating a certain power $P_2$, the transmit signal at U2 is given by $x_2 = \sqrt{P_2}s_2$. Thus, the signal received at the BS is given by
\begin{equation}
	y = h_1 x_1+ h_2 x_2+ z,
\end{equation}
where $h_i \ (i=1,2)$ denotes the channel coefficient of U$i$ and $z$ denotes the additive white Gaussian noise (AWGN) at the BS with zero-mean and variance $\sigma_n^2$. We assume that the decoding order is $s_{11} \rightarrow s_{2} \rightarrow s_{12}$. The BS first regards $s_{12}$ and $s_2$ as interference to decode $s_{11}$. Thus, the signal to interference noise ratio (SINR) of the first decoded stream $s_{11}$, denoted by $\gamma_{11}$, can be expressed as 
\begin{equation}
	\gamma_{11} = \frac{P_{11} G_1}{P_{12} G_1 + P_2 G_2 + \sigma_n^2},
\end{equation}
where $G_i = |h_i|^2 \ (i=1,2)$ denotes the channel gain of U$i$. Assuming that $s_{11}$ is successfully decoded, the BS removes $s_{11}$ and decodes $s_{2}$ while treating $s_{12}$ as noise. Through SIC, the SINR of the second decoded stream $s_2$, denoted by $\gamma_{22}$, can be expressed as 
\begin{equation}
	\gamma_{22} = \frac{P_2 G_2}{P_{12} G_1 + \sigma_n^2}.
\end{equation}
Next, the BS removes $s_{2}$ and decodes $s_{12}$ when $s_{2}$ is successfully decoded. The SINR of the third decoded stream $s_{12}$, denoted by $\gamma_{12}$, can be expressed as 
\begin{equation}
	\gamma_{12} = \frac{P_{12} G_1}{\sigma_n^2}.
\end{equation}

\section{Performance Analysis of Uplink RSMA\\ in The FBL Regime}\label{Sec:FBL}
\subsection{Blocklength Structure And Achievable Rate Region}
\begin{figure}[htbp]
	\centering\includegraphics[width=3.3in]{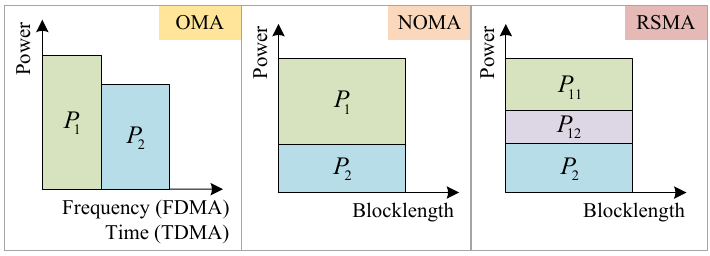}
	\caption{The power and blocklength of FDMA, TDMA, NOMA, and RSMA.}\label{fig:3MA}
\end{figure}
In the FBL regime, the blocklength is denoted by $n=TB$, where $T$ represents the time span (TTI) and $B$ represents the frequency resource occupied by the current block. As illustrated in Fig.~\ref{fig:3MA}, orthogonal multiple access (OMA), such as FDMA and TDMA, transmits signals in different frequency and time domains, i.e., different signals occupy different blocklengths~\cite{Blocklength_comprision}. As for NOMA and RSMA, they transmit signals in different power domains. Thus, NOMA and RSMA can share a common blocklength, which means the total required blocklength of RSMA and NOMA is smaller and leads to a lower delay when the bandwidth is fixed.

In the IBL regime, RSMA can reach every point of the Gaussian MAC capacity region with the error probability $\varepsilon \rightarrow 0$. Thus, the achievable rate region of uplink RSMA in the IBL regime is given by $R_1^{\rm IBL} = R_{11}^{\rm IBL}+R_{12}^{\rm IBL} \leq C\left(\gamma_1\right)$, $R_2^{\rm IBL} \leq C\left(\gamma_2\right)$, and $R_1^{\rm IBL} + R_{2}^{\rm IBL} \leq C(\gamma_{\rm sum})$, where $C (\gamma_i) = {\log_2}\left(1 + \gamma_{i}\right)\  \left(i = 1,2,\rm sum\right) $ denotes the Shannon capacity, $\gamma_{1} =\frac{P_{1} G_1}{P_2 G_2+\sigma_n^2}$, $\gamma_{2} =\frac{P_2 G_2}{P_{1} G_1+\sigma_n^2}$, $\gamma_{\rm sum} = \frac{P_1 G_1 + P_2 G_2}{\sigma_n^2}$, and $P_1 = P_{11}+P_{12}$. However, in the case of FBL regime, the error probability no longer approaches to $0$ and the blocklength has an impact on the achievable rate. To investigate the suitable RSMA scheme for short packet transmissions, we need to further analyze the impact of blocklength on the achievable rate and capacity region of uplink RSMA. The achievable rate of $s_i$ in the FBL regime, denoted by $R_{i}\left(n, \gamma_{i}\right)$, can be approximated as~\cite{FBL}
\begin{align}\label{Eq:Rate}
	R_{i}\left(n, \gamma_{i}\right) \approx {\log_2}\left(1 + \gamma_{i} \right) - \sqrt {\frac{V_{i}}{n}} Q^{-1}(\varepsilon_{i}){\log _2}e,\nonumber \\ 
	\hspace{3cm} i=\{11, 12, 22\},
\end{align}
where $V_{i}=1-(1+\gamma_{i})^{-2} $ and $\varepsilon_{i}$ denote the channel dispersion and the predefined error probability of stream $s_i\ (i =11,12,22)$, and $Q^{-1}(\cdot)$ denotes the inverse of Q-function. Thus, the achievable rate region of U1 and U2 in the FBL regime can be expressed as follows:
\begin{equation}
	\begin{cases} 
		R_1^{\rm FBL} \leq C\left(\gamma_1\right) -D_1,\\
		R_2^{\rm FBL} \leq C\left(\gamma_2\right) -D_2,\\
		R_1^{\rm FBL} + R_2^{\rm FBL} \\ 
		\hspace{0.9cm} \leq R_{11}\left(n, \gamma_{11}\right) + R_{12}\left(n, \gamma_{12}\right) + R_{22}\left(n, \gamma_{22}\right)\\ 
		\hspace{0.9cm} = C(\gamma_{\rm sum}) - D_{11} - D_{12}  - D_{22},
	\end{cases}
\end{equation}
where $D_{i}= \sqrt {\frac{1-(1+\gamma_{i})^{-2}}{n}} Q^{-1}(\varepsilon_{i}){\log_2}e\ (i = 1,2,11,12,22)$.

\emph{Proof:} The proof is provided in Appendix.

\subsection{Error Probability And Effective Throughput}
According to the decoding order, the error probability of $W_1$, denoted by $\varepsilon_1$, can be expressed as 
\begin{equation}
	\varepsilon_1 = \varepsilon_{11} + \left( 1- \varepsilon_{11} \right) \varepsilon_{22} + \left( 1- \varepsilon_{11} \right) \left( 1- \varepsilon_{22} \right) \varepsilon_{12}.
\end{equation}
Since the reliability requirement in URLLC is relatively small (e.g., $10^{-5} \sim 10^{-9}$), the product of two error probabilities can be omitted. Thus, the error probability of $W_1$ can be approximated as 
\begin{equation}
	\varepsilon_1 \approx \varepsilon_{11} + \varepsilon_{12} + \varepsilon_{22}.
\end{equation}
Similarly, the error probability of $W_2$, denoted by $\varepsilon_2$, can be expressed as 
\begin{equation}
	\varepsilon_2 = \varepsilon_{11} + \left( 1- \varepsilon_{11} \right) \varepsilon_{22} \approx \varepsilon_{11} + \varepsilon_{22}.
\end{equation}
Based on the above analysis, the effective throughput of U1 and U2, denoted by $T_{1}$ and $T_{2}$, can be given as follows:
\begin{equation}
	\begin{cases} 
	T_{1} = \left( 1 - \varepsilon_{1} \right) n  \big[R_{11}\left(n, \gamma_{11}\right)+R_{12}\left(n, \gamma_{12}\right)\big],\\
	T_{2} = \left( 1 - \varepsilon_{2} \right) n R_{2}\left(n, \gamma_{22}\right).
	\end{cases}
\end{equation}

\section{Uplink RSMA-Based Blocklength Minimization Problem}\label{Sec:Problem}
To satisfy stringent delay requirements in URLLC, we propose the uplink RSMA-based blocklength minimization problem with power allocation scheme for short packet transmissions under reliability and effective throughput demands.

\subsection{Problem Formulation}
In this paper, we aim to minimize the blocklength while meeting reliability and effective throughput requirements. Thus, the blocklength minimization problem, denoted by $\textbf{\textit{P}1}$, can be expressed as
\begin{subequations}
\begin{align}\label{P1}
	\textbf{\textit{P}1:\ } &\min_{  \boldsymbol P}\ n \\ 
	\  \mathrm{s.t.}\ \  & P_{11} + P_{12} \leq P_{\rm t}, \label{Eq:P1_Pt1}\\
	\ \  & P_2 \leq P_{\rm t}, \label{Eq:P1_Pt2}\\
	\ \  & T_i \geq T^{\rm th}_i, \ i = \{1,2\}, \label{Eq:P1_Tth}\\
	\ \  & N_{\min} \leq n \leq N_{\max}, \label{Eq:P1_Nmax}
\end{align}
\end{subequations}
where $\boldsymbol P = \left[ P_{11}, P_{12}, P_{2} \right]$, $P_{\rm t}$ denotes the maximum transmit power, $T^{\rm th}_i$ denotes the effective throughput threshold of U$i$, $N_{\min}$ and $N_{\max}$ denote the minimum and maximum blocklength. Constraint (\ref{Eq:P1_Tth}) guarantees the effective throughput demands, while the error probability of each stream is predefined to ensure the reliability requirements. The blocklength minimization problem $\textbf{\textit{P}1}$ is non-convex due to constraint (\ref{Eq:P1_Tth}) with coupled optimization variables. 
 
\subsection{Problem Transformation}
To deal with the non-convex problem $\textbf{\textit{P}1}$, we introduce slack variables $\boldsymbol{\delta} = \left[\delta_{11}, \delta_{12}, \delta_{22} \right]$ and $\boldsymbol{\tau} = \left[\tau_{11}, \tau_{12} \right]$, where $\boldsymbol\delta$ and $\boldsymbol\tau$ are the lower bounds of SINR and effective throughput, respectively. With the introduced slack variables, $\textbf{\textit{P}1}$ can be written as follows:
\begin{subequations}
	\begin{align}\label{P2}
		\textbf{\textit{P}2:\ } &\min_{  \boldsymbol P}\ n \\ 
		\  \mathrm{s.t.}\ \  & P_{11} + P_{12} \leq P_{\rm t}, \label{Eq:P_P_Pt_1}\\
		\ \  & P_{2} \leq P_{\rm t}, \label{Eq:P_P_Pt_2}\\
		\ \  & \left(1-\varepsilon_1\right) n \Big[\log_2\left(1+\delta_{1i}\right) - E_{1} \nu_{1i}\Big]  \nonumber \\ &\hspace{4.2cm}\geq \tau_{1i}, \ i=\{1,2\} \label{Eq:P_P_tau_1}\\
		\ \  & \left(1-\varepsilon_2\right) n \Big[\log_2\left(1+\delta_{22}\right)-E_2 \nu_{2}\Big] \geq T^{\rm th}_2, \label{Eq:P_P_tau_2}\\
		\ \  & \frac{P_{11} G_1}{P_{12} G_1 + P_2 G_2 + \sigma_n^2} \geq \delta_{11}, \label{Eq:P_P_delta_11}\\
		\ \  &  \frac{P_2 G_2}{P_{12} G_1 + \sigma_n^2} \geq \delta_{22}, \label{Eq:P_P_delta_22}\\
		\ \  & \frac{P_{12} G_1}{\sigma_n^2}\geq \delta_{12}, \label{Eq:P_P_delta_12}\\
		\ \  & \tau_{11}+\tau_{12}\geq T_{\rm th}, \label{Eq:P_P_Tth_1}
	\end{align}
\end{subequations}
where $E_{1} = \frac{Q^{-1}(\varepsilon_1)}{\sqrt{n}}{\log_2}e$, $E_{2} = \frac{Q^{-1}(\varepsilon_2)}{\sqrt{n}}{\log_2}e$, and $\nu_{1i} = \sqrt{1-\left(1+\delta_{1i}\right)^2}\ (i=1,2 )$. Due to the non-convexity of constraints (\ref{Eq:P_P_tau_1})-(\ref{Eq:P_P_delta_22}), $\textbf{\textit{P}2}$ is still non-convex. Thus, we use the first order Taylor series to approximate the non-convex part in the constraints. Constraints (\ref{Eq:P_P_tau_1}) and (\ref{Eq:P_P_tau_2}) can be approximated at the point $\boldsymbol \delta^{(t)}$ at the $t$-th iteration as follows:
\begin{subequations}
	\begin{align}\label{T_1_appro}
		&\left(1-{\varepsilon}_1\right) n  \Bigg[\log_2\left(1+\delta_{1i}\right)-E_{1} \Bigg\{ \left[1-\left(1+\delta_{1i}^{(t)}\right)^{-2}\right]^{\frac{1}{2}}  \nonumber\\ 
		&+ \left(\delta_{1i}-\delta_{1i}^{(t)}\right) \left[1-\left(1+\delta_{1i}^{(t)}\right)^{-2}\right]^{-\frac{1}{2}}
		\nonumber\\ 
		&\hspace{2cm} \times \left(1+\delta_{1i}^{(t)}\right)^{-3} \Bigg\}\Bigg] \geq \tau_{1i}, \ i=\{1,2\}, \\ 
		&\text{and} \nonumber\\
		&\left(1-{\varepsilon}_2\right) n  \Bigg[\log_2\left(1+\delta_{22}\right)-E_{2} \Bigg\{ \left[1-\left(1+\delta_{22}^{(t)}\right)^{-2}\right]^{\frac{1}{2}} + \nonumber\\ 
		&\left(\delta_{22}-\delta_{22}^{(t)}\right) \left[1-\left(1+\delta_{22}^{(t)}\right)^{-2}\right]^{-\frac{1}{2}} \left(1+\delta_{22}^{(t)}\right)^{-3} \Bigg\}\Bigg] \geq T^{\rm th}_2.\label{T_2_appro}
	\end{align}
\end{subequations}
Constraints (\ref{Eq:P_P_delta_11}) and (\ref{Eq:P_P_delta_22}) can be approximated at the point $\boldsymbol \delta^{(t)}$ and $\boldsymbol P^{(t)}$ at the $t$-th iteration as follows:
\begin{subequations}
	\begin{align}\label{SINR_1_appro}
		&\hspace{0.4cm}P_{12} G_1 + P_2 G_2 + \sigma_n^2 - \frac{P_{11} G_1}{\delta_{11}^{(t)}} \nonumber \\ &\hspace{3.5cm} + \left( \delta_{11} - \delta_{11}^{(t)}\right) \frac{P_{11}^{(t)} G_1}{\left(\delta_{11}^{(t)}\right)^2} \leq 0, \\ 
		&\text{and} \nonumber\\
		&\hspace{0.25cm}P_{12} G_1 + \sigma_n^2  -\frac{P_2 G_2}{\delta_{22}^{(t)}}  \ + \left( \delta_{22} - \delta_{22}^{(t)}\right) \frac{P_2^{(t)} G_2}{\left(\delta_{22}^{(t)}\right)^2} \leq 0.\label{SINR_2_appro} 
	\end{align}
\end{subequations}
Based on the above first order Taylor series approximations, the non-convex problem  $\textbf{\textit{P}2}$ can be transformed into a convex problem as follows:
	\begin{align}\label{P3}
		\textbf{\textit{P}3:\ } &\min_{\boldsymbol P, \boldsymbol \delta, \boldsymbol \tau}\ n \\ 
		\  \mathrm{s.t.}\ \  &\text{(\ref{Eq:P_P_Pt_1}),(\ref{Eq:P_P_Pt_2}),(\ref{Eq:P_P_delta_12}),(\ref{Eq:P_P_Tth_1}),}\text{(\ref{T_1_appro}),(\ref{T_2_appro}),(\ref{SINR_1_appro}),(\ref{SINR_2_appro})}.\nonumber
	\end{align}
All the constraints are transformed into convex, thus $\textbf{\textit{P}3}$ can be solved using the iterative alternating optimization (AO) method. At the $t$-th iteration, based on the optimal solution obtained from the $(t-1)$-th iteration $\left(\boldsymbol P^{(t-1)},\boldsymbol \delta^{(t-1)}, \boldsymbol \tau ^{(t-1)}\right)$, solving the convex problem $\textbf{\textit{P}3}$ to get the optimal solution of the $t$-th iteration. Through alternating iteration, the corresponding minimum blocklength can be obtained until the blocklength converges. The detailed iterative algorithm flow is outlined in Algorithm 1.
\begin{algorithm}[h]
	\caption{AO algorithm for solving $\textbf{\textit{P}1}$.}
	\begin{algorithmic}[1]
		\STATE  Initialization: Initialize the index of iteration $t=0$, the initial blocklength $n^{(0)} = 10^6$, the converge threshold $\xi$, and the transmit power $\boldsymbol P^{(0)}$. Initialize the slack variables $\boldsymbol \delta^{(0)}$ and $\boldsymbol \tau^{(0)}$.
		\REPEAT
		\STATE With the given $\boldsymbol P ^{(t)}$, $\boldsymbol \delta ^{(t)}$, and $\boldsymbol \tau ^{(t)}$, optimize $n^{(t+1)}$ by solving problem $\textbf{\textit{P}3}$ and get the optimal solution $\boldsymbol P^{(t)*}$, $\boldsymbol \delta ^{(t)*}$, and $\boldsymbol \tau ^{(t)*}$.
		\STATE Update $\boldsymbol P ^{(t+1)}$, $\boldsymbol \delta ^{(t+1)}$, and $\boldsymbol \tau ^{(t+1)}$ as $\boldsymbol P^{(t)*}$, $\boldsymbol \delta ^{(t)*}$, and $\boldsymbol \tau ^{(t)*}$.
		\STATE Update $t = t+1$.
		\UNTIL {$\left|n^{(t)} - n^{(t+1)}\right| \leq \xi$.}
	\end{algorithmic}
\end{algorithm}

\subsection{Comparison to NOMA}
In order to perform a comparative analysis with uplink RSMA, here we provide the performance analysis for two-user uplink NOMA, FDMA, and TDMA networks.

In uplink NOMA, the messages $W_{1}$ and $W_{2}$ of U1 and U2 are encoded into streams $s_{1}$ and $s_{2}$. The BS first decodes $s_1$ of the strongest user U1 while treating $s_2$ as interference. After decoding $s_1$, the BS removes it and decodes $s_2$. Thus, the SINR of $s_{1}$ is $\gamma_{11}^{\rm N} = \frac{P_1 G_1}{P_2 G_2 + \sigma_n^2}$ and the SINR of $s_{2}$ is $\gamma_{22}^{\rm N} = \frac{P_2 G_2}{\sigma_n^2}$. Based on the SINRs and (\ref{Eq:Rate}), we can get the achievable rate $R_1^{\rm N}\left(n, \gamma_{11}^{\rm N}\right)$ and $R_{2}^{\rm N}\left(n, \gamma_{22}^{\rm N}\right)$ of $s_{1}$ and $s_{2}$ in NOMA. The corresponding error probability of each stream in NOMA is set as $\varepsilon_{i}^{\rm N}\ (i=11,22)$. According to NOMA decoding order, the error probability of $W_1$ is $\varepsilon_1^{\rm N} = \varepsilon_{11}^{\rm N} + \left( 1- \varepsilon_{11}^{\rm N} \right) \varepsilon_{22}^{\rm N} \approx \varepsilon_{11}^{\rm N} + \varepsilon_{22}^{\rm N}$, and the error probability of $W_2$ is $\varepsilon_2^{\rm N} = \varepsilon_{22}^{\rm N}$. Therefore, the effective throughput of U1 in uplink NOMA is $T_{1}^{\rm N} = \left( 1 - \varepsilon_{1}^{\rm N} \right) n R_{1}^{\rm N} \left(n, \gamma_{11}^{\rm N}\right)$ and the effective throughput of U2 in uplink NOMA is
$T_{2}^{\rm N} = \left( 1 - \varepsilon_{2}^{\rm N} \right) n R_{2}^{\rm N} \left(n, \gamma_{22}^{\rm N}\right)$. Based on the above analysis, the blocklength minimization problem for uplink NOMA can be expressed as follows:
\begin{align}\label{P4}
	\textbf{\textit{P}4:\ } &\min_{  \boldsymbol P^{\rm N}}\ n \\ 
	\  \mathrm{s.t.}\ \  &P_1 \leq P_{\rm t},\text{(\ref{Eq:P1_Pt2}),(\ref{Eq:P1_Tth}),(\ref{Eq:P1_Nmax})},\nonumber
\end{align}
where $\boldsymbol P^{\rm N}=\left[ P_{1}, P_{2} \right] $.
\subsection{Comparison to OMA}
We assume the bandwidth fraction is $\alpha^{\rm F}$ in uplink FDMA and the time fraction is $\alpha^{\rm T}$ in uplink TDMA. Thus, the blocklengths of U1 and U2 are $n_1^{\rm F} =\alpha^{\rm F} n $ and $n_2^{\rm F}=(1-\alpha^{\rm F})n$ in FDMA, while $n_1^{\rm T} =\alpha^{\rm T} n $ and $n_2^{\rm T}=(1-\alpha^{\rm T})n$ in TDMA, respectively. As a result, the SINRs of FDMA are given by $\gamma_{1}^{\rm F} = \frac{P_1 G_1}{\alpha^{\rm F} \sigma_n^2}$ and $\gamma_{2}^{\rm F} = \frac{P_2 G_2}{\left(1-\alpha^{\rm F}\right) \sigma_n^2}$, while the SINRs of TDMA are given by $\gamma_{1}^{\rm T} = \frac{P_1 G_1}{\sigma_n^2}$ and $\gamma_{2}^{\rm T} = \frac{P_2 G_2}{\sigma_n^2}$, respectively. The corresponding error probabilities are given by $\varepsilon_{i}^{j}\ (i=\{1,2\},j=\{\rm{F,T}\})$. Then, the achievable rate of U1 is $R_{1}^{j} \left(n_1^{j}, \gamma_{1}^{j}\right) \approx \alpha^j {\log_2}\left(1 + \gamma_{1}^{j} \right) - \sqrt {\frac{V_1^{j}}{n_1^{j}}} Q^{-1}(\varepsilon_1^{j}){\log_2}e$ when $j={\rm F}$ for FDMA and $j={\rm T}$ for TDMA. Similarly, the achievable rate of U2 is $R_{2}^{j} \left(n_2^{j}, \gamma_{2}^{j}\right) \approx \left(1-\alpha^{j} \right) {\log_2}\left(1 + \gamma_{2}^{j} \right) - \sqrt {\frac{V_2^{j}}{n_2^{j}}} Q^{-1}(\varepsilon_2^{j}){\log_2}e$ when $j={\rm F}$ for FDMA and $j={\rm T}$ for TDMA. Thus, the effective throughput is $T_{i}^{j} = \left( 1 - \varepsilon_{i}^{j} \right) n_i^j R_{i}^{j} \left(n_i^j, \gamma_{i}^{j}\right)\ \left(i=\{1,2\},j=\{\rm{F,T}\}\right)$. Therefore, the blocklength minimization problem for OMA ($j=\rm F$ for FDMA and $j=\rm T$ for TDMA) can be expressed as follows:
\begin{align}\label{P5}
	\textbf{\textit{P}5:\ } &\min_{  \boldsymbol P^j}\ n_1^j + n_2^j \\ 
	\  \mathrm{s.t.}\ \  &P_1 \leq P_{\rm t},\text{(\ref{Eq:P1_Pt2}),(\ref{Eq:P1_Tth}),(\ref{Eq:P1_Nmax})},\nonumber
\end{align}
where $\boldsymbol P^j=\left[ P_{1}, P_{2} \right]\ \left(j=\{\rm{F,T}\}\right) $.

\section{Numerical Results}\label{Sec:Results}
In this section, we evaluate the achievable rate region performance and the proposed blocklength minimization of uplink RSMA. We set the channel gain $G_1 = 1$ for U1 and $G_2 = 0.7$ for U2 with $G_1>G_2$. Without loss of generality, we assume the noise variance $\sigma^2=1$. For NOMA, we set NOMA-12 with the decoding order $s_1 \rightarrow s_2$ and NOMA-21 with the decoding order $s_2 \rightarrow s_1$. We set the bandwidth fraction and time fraction as $\alpha^{\rm F} = \alpha^{\rm T} = \frac{P_1}{P_1+P_2}$ in FDMA and TDMA.  To satisfy the reliability requirements in URLLC, the predefined error probability of each stream is set to $\varepsilon = 10^{-6}$ to ensure the overall error probability of uplink RSMA is lower than $10^{-5}$. In addition, we set the maximum blocklength $N_{\max}=3000$ and the minimum blocklength $N_{\min} = 100$~\cite{minimum_blocklength}.

\begin{figure}[htbp]
	\centering\includegraphics[width=3in]{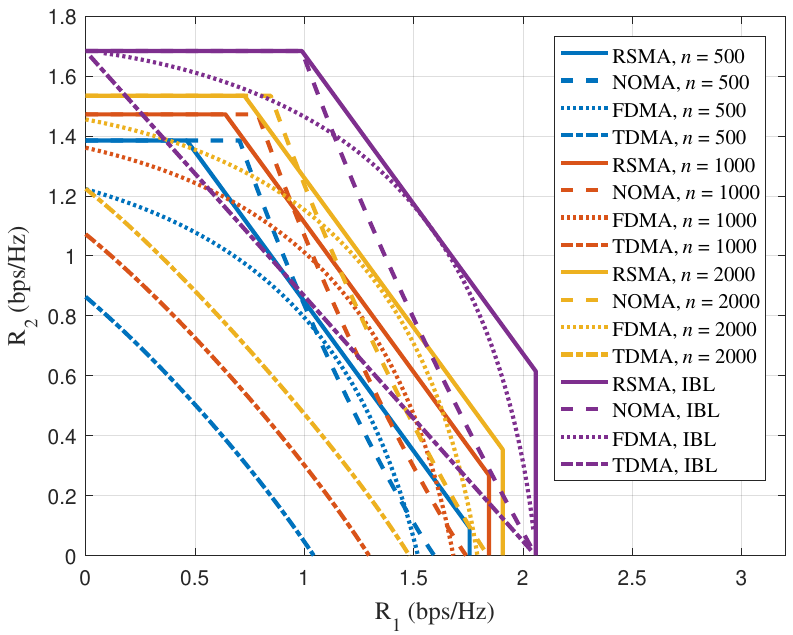}
	\caption{Achievable rate region comparison of different MA strategies with different blocklengths.}\label{fig:capacity_region}
\end{figure}
Figure~\ref{fig:capacity_region} depicts the achievable rate region comparison of different MA strategies with different blocklengths as $n$ set to $500, 1000, 2000$, and infinite. The achievable rates of U1 and U2 both increase as blocklength increases, and therefore the achievable rate region also expands with the increase of blocklength. In the traditional IBL regime, uplink RSMA can achieve the Gaussian MAC capacity region, while NOMA without time sharing can achieve maximum rate only at one of the users. This is because by changing the transmit power allocation between two data streams $s_{11}$ and $s_{12}$, RSMA can bridge NOMA-12 (allocate all the transmit power to $s_{11}$) and NOMA-21 (allocate all the transmit power to $s_{12}$), which leads to a larger rate region without time sharing. FDMA can only reach the capacity region at one point with $\alpha^{\rm F} = \frac{P_1}{P_1+P_2}$, while TDMA cannot reach the capacity region without variable transmit power. 

However, in the FBL regime, uplink RSMA cannot achieve the Gaussian MAC capacity region and the achievable rate region of RSMA is not always larger than that of NOMA. On the one hand, the error probability no longer approaches $0$ with the FBL code, resulting in a decrease in the achievable rate. On the other hand, the signal $s_1$ is divided into two streams in uplink RSMA, which brings more channel dispersion terms than NOMA, resulting in a decrease in the total achievable rate. According to $D = \sqrt {\frac{V}{n}} Q^{-1}(\varepsilon){\log_2}e$, it can be observed that as the blocklength increases, the impact of channel dispersion gradually decreases. As shown in Fig.~\ref{fig:capacity_region}, the rate region of RSMA can gradually include NOMA with blocklength increasing.

\begin{figure}[htbp]
	\begin{center}
		\subfigure[Blocklength versus the maximum transmit power with $T^{\rm th}_1>T^{\rm th}_2$.]{
			\includegraphics[width=1.61in]{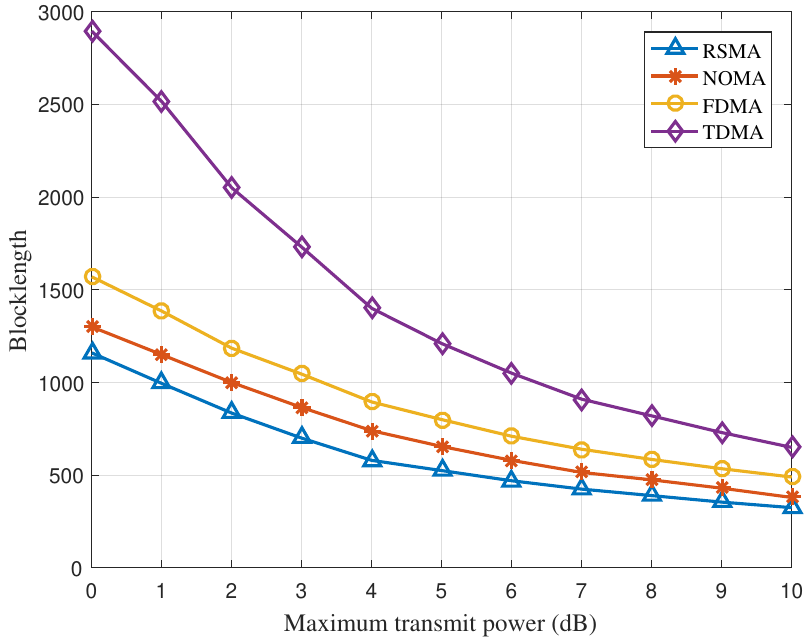}\label{fig:n_Pt}
		}
		\subfigure[Blocklength versus the maximum transmit power with $T^{\rm th}_1<T^{\rm th}_2$.]{
			\includegraphics[width=1.61in]{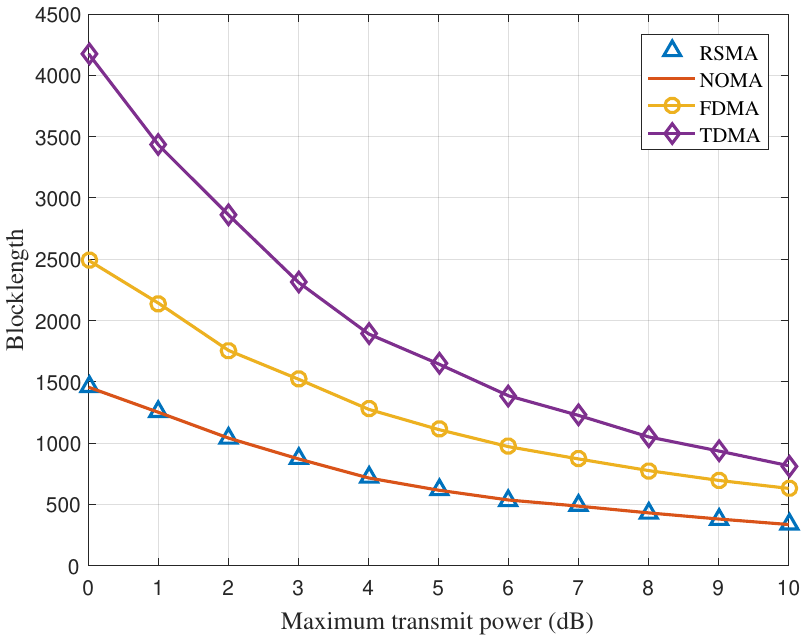}\label{fig:n_Pt_T1_200}
		}
		\caption{Blocklength comparison of different MA strategies versus the maximum transmit power.}
	\end{center}
\end{figure}
Figure~4 shows the blocklength comparison of different MA strategies versus the maximum transmit power with different effective throughput requirements. The blocklength gradually decreases as the maximum transmit power increases. As shown in Fig.~\ref{fig:n_Pt}, although RSMA and NOMA can use the common blocklength without dividing it to two users, RSMA can achieve a smaller blocklength. This is because NOMA decodes $s_1$ first and removes $s_1$ when decoding $s_2$, which can guarantee a high rate of $s_2$ but result in a relatively low rate of $s_1$. When $T^{\rm th}_1$ is large, U1 needs a larger blocklength to reach $T^{\rm th}_1$ than that of RSMA. In FDMA and TDMA, U1 and U2 cannot share the common blocklength. As a result, the blocklength is divided into U1 and U2 in the frequency domain and time domain, resulting in a relatively large blocklength. It can be seen from Fig.~\ref{fig:n_Pt_T1_200} that the blocklengths of RSMA and NOMA are the same, which means NOMA can achieve the same rate as RSMA with the same blocklength. This is because when the effective throughput requirement is set as $T^{\rm th}_1<T^{\rm th}_2$, the guaranteed rate of $s_2$ can satisfy a larger requirement $T^{\rm th}_ 2$. However, RSMA can adapt to different throughput requirements and achieve the smallest blocklength, which means it is more suitable and flexible than NOMA for heterogeneous networks with different QoS requirements.

\begin{figure}[htbp]
	\begin{minipage}[t]{0.48\linewidth}
		\centering \label{fig:n_L}
		\includegraphics[width=1.85in]{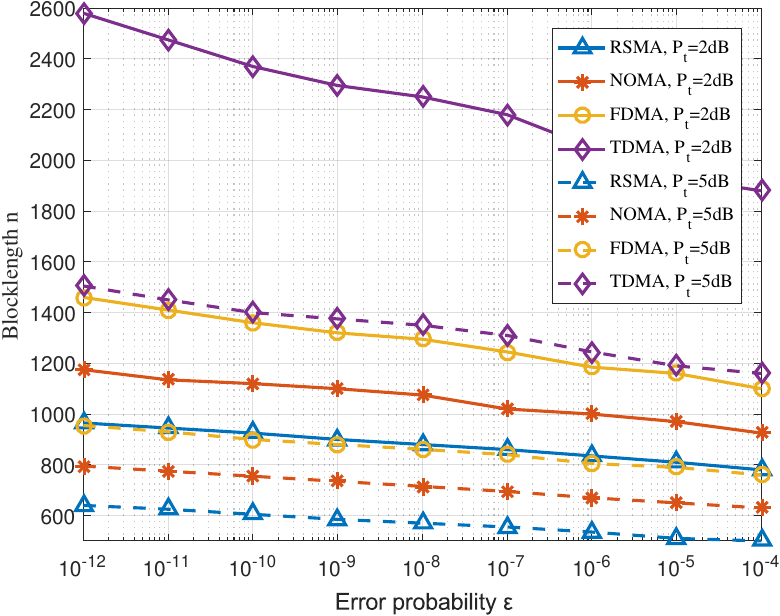}
		\caption{The blocklength comparison of different MA strategies versus the error probability with $P_t = 2$~dB and $P_t = 5$~dB.}
	\end{minipage}
	\hfill
	\begin{minipage}[t]{0.48\linewidth}	
		\centering \label{fig:R_n}
		\includegraphics[width=1.75in]{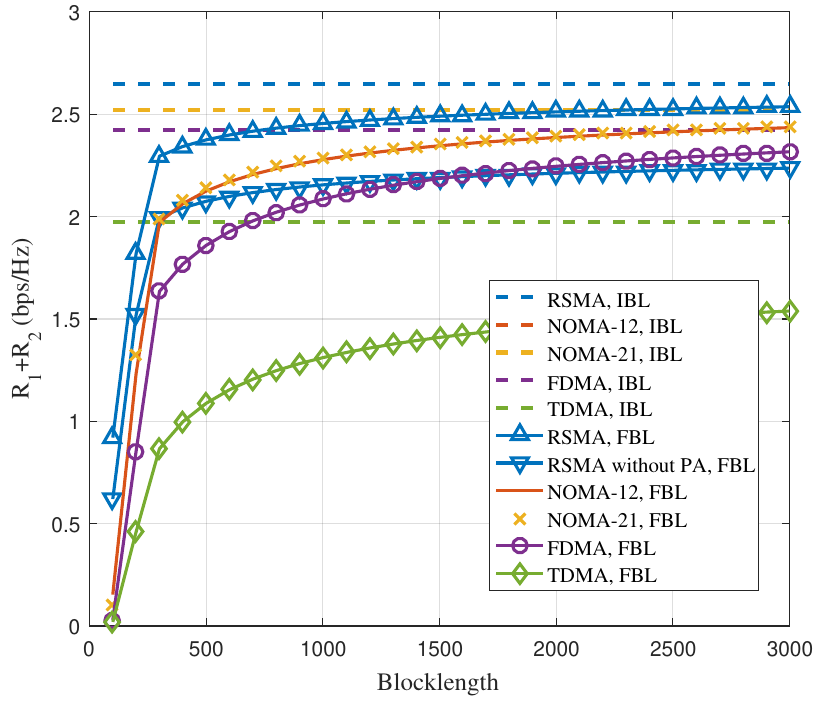}
		\caption{The sum of achievable rates comparison of different MA strategies versus the blocklength with $P_t = 5$~dB.}	
	\end{minipage}
\end{figure}

Figure~5 plots the blocklength comparison of different MA strategies versus the error probability with different maximum transmit powers. The blocklength decreases as the error probability increases for the four MA strategies, while the blocklength of RSMA is always lower than that of NOMA, FDMA, and TDMA. In addition, for vertical businesses with different reliability requirements in URLLC, as long as their reliability requirements are met, the minimum blocklength can be selected to achieve the minimum delay. 

Figure~6 shows the sum of the achievable rates versus the blocklength. RSMA can achieve the highest rate in both IBL and FBL regimes. In the FBL regime, the sum of achievable rates increases with the blocklength increasing. Therefore, the minimum blocklength can be selected as long as the rate requirements are achieved to satisfy low-delay requirements in URLLC. In addition, for a fixed value of blocklength, the achievable rate of RSMA without power allocation (PA) is lower than those of RSMA, NOMA-12, and NOMA-21 with power allocation. This means that through power allocation can RSMA achieve the function of bridging NOMA-12 and NOMA-21, resulting in a higher rate with the same blocklength.

\section{Conclusion}\label{Sec:Con}
In this paper, we solved the problem of minimizing the blocklength with the power allocation for uplink RSMA in URLLC. In particular, we analyze the performance of uplink RSMA in the FBL regime, in terms of the achievable rate region and the effective throughput. On this basis, we proposed the uplink RSMA-based blocklength minimization problem under the reliability and effective throughput constraints to satisfy the low-delay and reliability requirements in URLLC. Furthermore, we developed an alternating optimization algorithm to solve this non-convex problem to obtain the optimal power allocation and the minimum blocklength. Numerical results demonstrated that uplink RSMA cannot achieve the Gaussian MAC capacity region in the FBL regime. However, with the help of our proposed blocklength minimization scheme, uplink RSMA can significantly reduce the blocklength to achieve a lower delay compared to uplink NOMA, FDMA, and TDMA, showing the potential of uplink RSMA for URLLC.

\appendix
\renewcommand{\appendixname}{Appendix~\Alph{section}}
The achievable rate of U1 can be expressed as follows:
	\begin{align}
		& \hspace{0.5cm} R_{11} \left(n, \gamma_{11}\right) + R_{12} \left(n, \gamma_{12}\right) \nonumber \\
		& \approx {\log_2}\left(1 + \gamma_{11} \right) - D_{11} + {\log_2}\left(1 + \gamma_{12} \right) - D_{12} \nonumber\\
		& = \log_2 \left(1 + \frac{P_{11} G_1}{P_{12} G_1 + P_2 G_2 + \sigma_n^2} \right) - D_{11} \nonumber\\
		& \hspace{4cm} + \log_2 \left(1 + \frac{P_{12} G_1}{\sigma_n^2} \right) - D_{12} \nonumber\\
		& = \log_2 \bigg[\left(\frac{P_{11} G_1 + P_{12} G_1 + P_2 G_2 + \sigma_n^2}{P_{12} G_1 + P_2 G_2 + \sigma_n^2}\right) \nonumber\\
		&\hspace{2.8cm} \cdot \left(\frac{P_{12} G_1 + \sigma_n^2}{\sigma_n^2}\right) \bigg] - D_{11} - D_{12} \nonumber\\
		& = \log_2 \left[\frac{\left(\frac{P_{1} G_1 + P_2 G_2 + \sigma_n^2}{\sigma_n^2}\right)}{\left(\frac{P_{12} G_1 + P_{2} G_2 + \sigma_n^2}{P_{12} G_1 + \sigma_n^2}\right)}\right] - D_{11} - D_{12} \nonumber\\
		& = \log_2 \left(1+\frac{P_{1} G_1 + P_2 G_2}{\sigma_n^2}\right) \nonumber\\
		& \hspace{1.9cm} - \log_2 \left(1+\frac{P_2 G_2}{P_{12} G_1 + \sigma_n^2}\right) - D_{11} - D_{12}\nonumber\\
		& = C(\gamma_{\rm sum}) - R_{2}\left(n, \gamma_{22}\right) - D_{11} - D_{12}  - D_{22}.
	\end{align}
Thus, we have $R_{11} \left(n, \gamma_{11}\right) + R_{12} \left(n, \gamma_{12}\right) + R_{2}\left(n, \gamma_{22}\right) = C(\gamma_{\rm sum}) - D_{11} - D_{12}  - D_{22}$.

\section*{Acknowledgment}
This work was supported in part by the Key Area R\&D Program of Guangdong Province under Grant 2020B0101110003, in part by the National Key R\&D Program of China under Grant 2021YFC3002102, and in part by the Key R\&D Plan of Shaanxi Province under Grant 2022ZDLGY05-09.

\bibliographystyle{IEEEtran}
\bibliography{References-GC}
\end{document}